\documentclass[11pt]{article}
\usepackage{moriond,epsfig,graphics}

\def\be{\begin{equation}}
\def\ee{\end{equation}}
\def\bea{\begin{eqnarray}}
\def\eea{\end{eqnarray}}

\begin{document}
\vspace*{4cm}
\title{MORIOND ELECTROWEAK 2006: THEORY SUMMARY}

\author{JOSEPH D. LYKKEN}

\address{Theoretical Physics Department\\ 
Fermi National Accelerator Laboratory\\
P.O. Box 500, Batavia, IL 60510 USA}

\maketitle\abstracts{A concise look at the big picture of
particle physics, including the status of the
Standard Model, neutrinos, supersymmetry, extra
dimensions and cosmology. Based upon the theoretical
summary presented at the XLIst Rencontres de Moriond on
Electroweak Interactions and Unified Theories,
La Thuile, 11-18 March 2006.
}

\section{Dancing in the Dark}
\begin{center}
``It's promising for the future. The important matches are coming.''
\end{center}
\hspace*{9cm}\emph{-- Zin\'edine Zidane}

The irony and frustration of particle physics today is that
simultaneously we know so much and so little. The Standard Model (SM)
has successfully predicted hundreds of new phenomena, observed
and confirmed in detail by a triumphant progression of experiments.
Our understanding and technical grasp of the Standard Model has
matured greatly, mostly as a result of the intimate interaction
of theory with experiment. This mature understanding has guided
our speculative frameworks for physics Beyond the Standard Model
(BSM). These frameworks are well-motivated. They have impressive
explanatory potential for fundamental mysteries not addressed by
the Standard Model. Most BSM frameworks make dramatic predictions 
for discoveries at the Large Hadron Collider (LHC). 

Yet, as of the middle of 2006, we are very much in the dark as to  
what is the larger picture that subsumes the Standard Model.
Dozens of experiments, which could have yielded clear signals of
new physics, have instead shown only Standard Model physics to within
ever--shrinking error bars. Some recent discoveries are clearly not
explained by the SM: neutrino oscillations, dark matter, and the 
accelerating expansion of the universe. But these discoveries,
singly or collectively, do not point unambiguously to their BSM
resolution. 

\section{Discovering the Standard Model}

The Standard Model predicts a wealth of new particle physics phenomena.
It is misleading to refer to experimental observations of these phenomena
as ``tests'' of the Standard Model. The first observation of a new
particle, particle property, interaction, decay or symmetry violation
is a discovery, not a test. Detailed measurements of new phenomena 
have the potential to reveal discrepancies with the Standard Model,
but just as importantly they force us to develop a deeper and more
concrete understanding of the Standard Model itself.

\subsection{Top physics}
Experiments in Tevatron Run I discovered a
new strongly pair-produced heavy state which decays promptly 
to a $W$ and a $b$-jet. Now experiments in Run II are discovering
the properties of this new particle, how it fits into the Standard Model,
and clues to its potentially unique role in particle physics.
During the past year, the charge, spin, and $V-A$ coupling of this
particle have been directly measured, confirming its identity as
top \cite{Klute:2006tb}. 

The mass of top is now measured with
an accuracy of $\pm 1.5$\% in a single experiment \cite{Brubaker:2006wv}.
This has two important consequences. First, we can extract with
greater precision the virtual top contributions to electoweak radiative
corrections, giving us greater sensitivity to the radiative effects
of new physics, including the Higgs. Second, we have discovered
that the mass of the top quark, in its natural units of
$v/\sqrt{2}$, is equal to one: $\lambda_t =0.99 \pm 0.01$. This is
a striking result, certainly as striking as the much ballyhooed
unification of the SM gauge couplings, which requires the additional
assumption of TeV supersymmetry to achieve similar precision.
As far as I know, no theoretical model has been suggested which can
explain this fact, which is made even more bizarre by the large
hierarchies of SM fermion masses in general.

\subsection{B physics}

Results from BaBar and Belle continue to have a great impact on
at least three fronts:
\begin{itemize}
\item First observations of processes which could have
${\cal O}(1)$ contributions from new physics.
\item Over-constraining the CKM model of flavor and $CP$ violation.
\item Challenging (and thus improving) our understanding of
strong interaction and heavy quark physics, including lattice QCD.
\end{itemize}  
Progress along these three fronts is correlated. The prediction
of B decays, for example, requires a combination of electroweak physics,
perturbative QCD, and nonperturbative QCD, further complicated
by multiple scales. Similar challenges occur in the charm and
kaon sectors \cite{Gerard:2006tu}.
Exclusive decays involve hadronic form factors
which are estimated using unquenched lattice QCD. 
As reported \cite{Bozzi:2006wu,Ikado:2006fc} at Moriond,
lattice uncertainities for exclusive B and charm decays are now on the
order of 10\%, as cross-checked in the data itself. These are
expected to improve to 5\%, and in some cases the order 1\% accuracy
which is already obtained in kaon sector \cite{Palutan:2006su}.

For the exclusive hadronic decays $B\to K\pi$ and $B\to \pi\pi$,
$SU(3)$ isospin relations can be used to make consistency checks
in which hadronic uncertainties are minimized. The first step is
to use an isospin analysis of the various $B\to \pi\pi$ modes
to extract some hadronic parameters from data. Then $SU(3)$
isospin, with known factorizable $SU(3)$-breaking corrections,
is used to make predictions for ratios of $B\to K\pi$ rates.
For one such ratio the agreement between prediction and data
is good. For another ratio, one which is sensitive to electroweak
penguins, the agreement is poor. This ``$K-\pi$'' puzzle could
be a hint of new physics in the aforementioned electroweak 
penguins \cite{Buras:2003dj}
or it may reflect a misunderstanding of QCD. An improved analysis
announced at Moriond still finds a discrepancy \cite{Malcles:2006jf}.

The basic theory of inclusive B decays is an effective hamiltonian
approach \cite{Fleischer:2005vz} using an operator product expansion:
\bea
\langle f|H_{eff}|i\rangle = \frac{G_F}{\sqrt{2}}
\lambda_{CKM}\sum_k C_k(\mu )
\langle f|Q_k(\mu)|i\rangle
\; ,
\eea
where the Wilson coefficients $C_k(\mu ,\alpha_s)$ are the scale-dependent
couplings of the interactions induced by the operators $Q_k$. Higher order
operators are suppressed by powers of the small quantity
$\Lambda_{QCD}/m_b$. Recent successes of this method 
include \cite{Schietinger:2006rc,Gambino:2004mv}
the determination of the sign of $C_7$ in the inclusive rare
decays $b\to s\ell^+\ell^-$.

However our theoretical handle on inclusive B decays is still far
from satisfactory. For example, inclusive $B\to X_s \gamma$ decays
are among the most sensitive channels for supersymmetry and other
new physics. The NLO theoretical prediction for the inclusive
branching fraction has about a 10\% uncertainty, which is comparable
to the current experimental uncertainty \cite{Schietinger:2006rc}.
To compare with the future data sets, we need a NNLO calulation,
reducing the renormalization scheme-dependence on the
(poorly measured) charm quark mass, and dealing with the effects
of the 1 GeV scale $m_b - 2E_{min}^{\gamma}$ created by the
analysis cuts \cite{Becher:2006qw}.

The situation is especially challenging for the inclusive
semileptonic decays used to extract the CKM element $V_{ub}$.
Here the kinimatic cuts used to separate $b\to u$ from $b\to c$
cause a breakdown of the operator product expansion, which 
is patched up by introducing ``shape functions'' to resum
nonperturbative physics. It is thus not surprising that in
the latest global fits \cite{Bona:2006ah}
of B physics data to the Standard Model,
the largest inconsistency seems to come from the inclusive
determination of $|V_{ub}|$. This may indicate a problem either
with the central value (too large) or the estimated errors
(too small).

The most dramatic moment of the Moriond conference was the surprise
announcement by the DZero collaboration of the first two-sided
bound on the $B_s^0-\bar{B}_s^0$ mass difference $\Delta M_s$
\cite{Abazov:2006dm}. This was followed shortly after the conference
by a CDF measurement \cite{CDFBs}
with remarkable 2\% accuracy:
\bea
\Delta M_s({\rm CDF}) = 17.33^{+0.42}_{-0.21}({\rm stat.})\pm 0.07 
({\rm syst.}) {\rm\ ps}^{-1} \; .
\eea

It is exciting to observe this variety of matter-antimatter
oscillations never before seen. After Moriond a number of global
data fits, using also various lattice QCD inputs, have
attempted to estimate the Standard Model prediction for
$\Delta M_s$. The latest fit \cite{Bona:2006ah} gives:
\bea
\Delta M_s = 20.9 \pm 2.6 {\rm\ ps}^{-1} \; .
\eea
If we remove the problematic inclusive $|V_{ub}|$
determination from the analysis, the prediction becomes
\bea
\Delta M_s = 19.4 \pm 2.5 {\rm\ ps}^{-1} \; .
\eea
This is good agreement, but it is embarrassing that the
(data-assisted) theory error is 5 times the CDF experimental
error!

The CDF and DZero analyses
also provide strong constraints on new physics. Together
with the first branching ratio for the rare decay $B \to \tau \nu$,
reported by Belle shortly after Moriond \cite{Ikado:2006fc}, 
this follows a pattern
of first observations of the dwindling number of channels
in which ${\cal O}(1)$ signals of new physics could 
have been hiding.

To understand the significance of measuring $\Delta M_s$,
we first classify models of new physics according to 
how they affect flavor physics:
\begin{itemize}
\item {\bf CMFV:} These are models in which the only
source of quark flavor violation is the CKM matrix,
and the only low dimension operators contributing to
flavor transitions are those present already in the SM.
This is called Constrained Minimal Flavor Violation \cite{Buras:2003jf}.
Examples include minimal supergravity models with low or moderate
tan$\,\beta$, and models with a universal large extra dimension.
\item {\bf MFV:} Same as above, except there are some new
relevant operators. This is called Minimal 
Flavor Violation \cite{D'Ambrosio:2002ex}.
Examples include SUSY models with large
tan$\,\beta$, where the new relevant operators are Higgs penguins.
\item {\bf NMFV:} There are new operators involving the
third generation quarks, and these are flavor-diagonal up to
small rotations with roughly the same hierarchies as in
the CKM matrix. This is Next-to-Minimal Flavor Violation \cite{Agashe:2005hk}.
The quasi-alignment is an attractive way to solve the flavor
problems that appear in frameworks such as Little Higgs, topcolor,
and warped extra dimensions.
\item {\bf GFV:} These are models with both new operators and new
sources of flavor violation. This is called General Flavor
Violation \cite{Foster:2005kb}. Examples include most of the MSSM
parameter space, and almost any BSM model that you can think of,
before you start worrying about flavor constraints.
\end{itemize}

CMFV models predict the same relations among flavor parameters as
the SM \cite{Blanke:2006ig}, 
thus the CMFV prediction for $\Delta M_s$ is the same
as the SM prediction quoted above.
For MFV, the most interesting case is SUSY with large tan$\,\beta$.
Double Higgs penguins interfere destructively with the SM contribution,
reducing $\Delta M_s$ by an amount which is potentially enhanced
by (tan$\,\beta$/$M_A)^4$. At the same time, the rare decay
$B_s \to \mu^+\mu^-$ is enhanced by as much as (tan$\,\beta$/$M_A)^6$.
As it happens, the constraints on such models from $\Delta M_s$
are comparable at the moment to those from the CDF and DZero
upper limits \cite{Abulencia:2005pw,Abazov:2004dj}
on $B_s \to \mu^+\mu^-$. An improved CDF limit on this branching
fraction was announced at the time of Moriond:
\bea
Br(B_s \to \mu^+\mu^-) < 8 \times 10^{-8} \; (90\%{\rm\ CL}),
\quad < 1.0 \times 10^{-7} \; (95\%{\rm\ CL})\; .
\eea

\begin{figure}
\centerline{\includegraphics[width=.6\columnwidth]{carena.epsf}}
\caption{MFV SUSY models outside of the thick solid lines
are excluded. From hep-ph/0603106.}
\label{carena}
\end{figure}

Figure 1 shows the predictions \cite{Carena:2006ai}
for $Br(B_s \to \mu^+\mu^-)$
and the (negative) contribution to $\Delta M_s$, for a sample
of the large tan$\,\beta$ MFV SUSY parameter space. Obviously
the model space is beginning to be nontrivially constrained.
Better constraints are expected from improvement in the Tevatron
$Br(B_s \to \mu^+\mu^-)$ upper bound, to $2 \times 10^{-8}$
or lower. It has also been suggested \cite{Isidori:2006qy}
that the slightly high SM value for $\Delta M_s$, as well as the slightly
high SM value for $Br(B_u\to \tau\nu)$ compared to the Belle
result, are emerging signals of large tan$\,\beta$ SUSY.
If so, Tevatron and LHC results will provide definitive confirmation. 

As seen in Figure 2, NMFV models are significantly contrained by
the measurement of $\Delta M_s$, but ${\cal O}(1)$ new sources of
flavor violation are still allowed \cite{Ligeti:2006pm}.
LHCb will provide the definitive probe of these models, by
measuring (among other things) the time dependent $CP$ asymmetries
in $B_s^0$ decays.

For GFV models, the new results on $\Delta M_s$ are quite constraining
on a large piece of the general parameter space \cite{Foster:2006ze}.
In some cases
the $\Delta M_s$ constraint, even though it includes the large
theory error bar from the SM prediction, is much more constraining
than the current upper bound on $Br(B_s \to \mu^+\mu^-)$.
This is shown in Figure 3. It is interesting to note that the
preliminary versions of these plots were produced during the week
of Moriond, showing once again the remarkably tight coupling between theory
and experiment in this field.

\begin{figure}
\centerline{\includegraphics[width=.4\columnwidth]{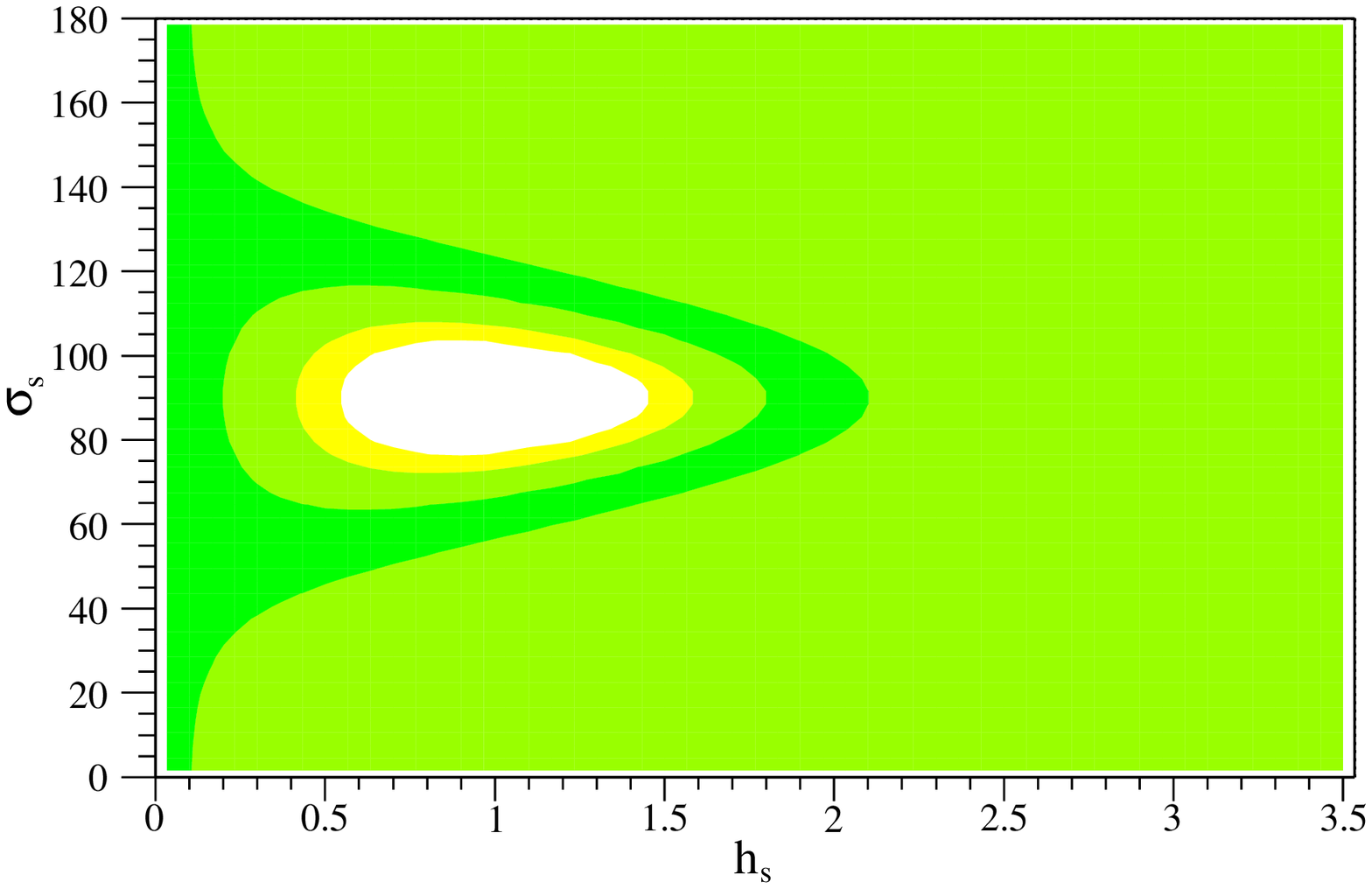} \hfil
  \includegraphics[width=.4\columnwidth]{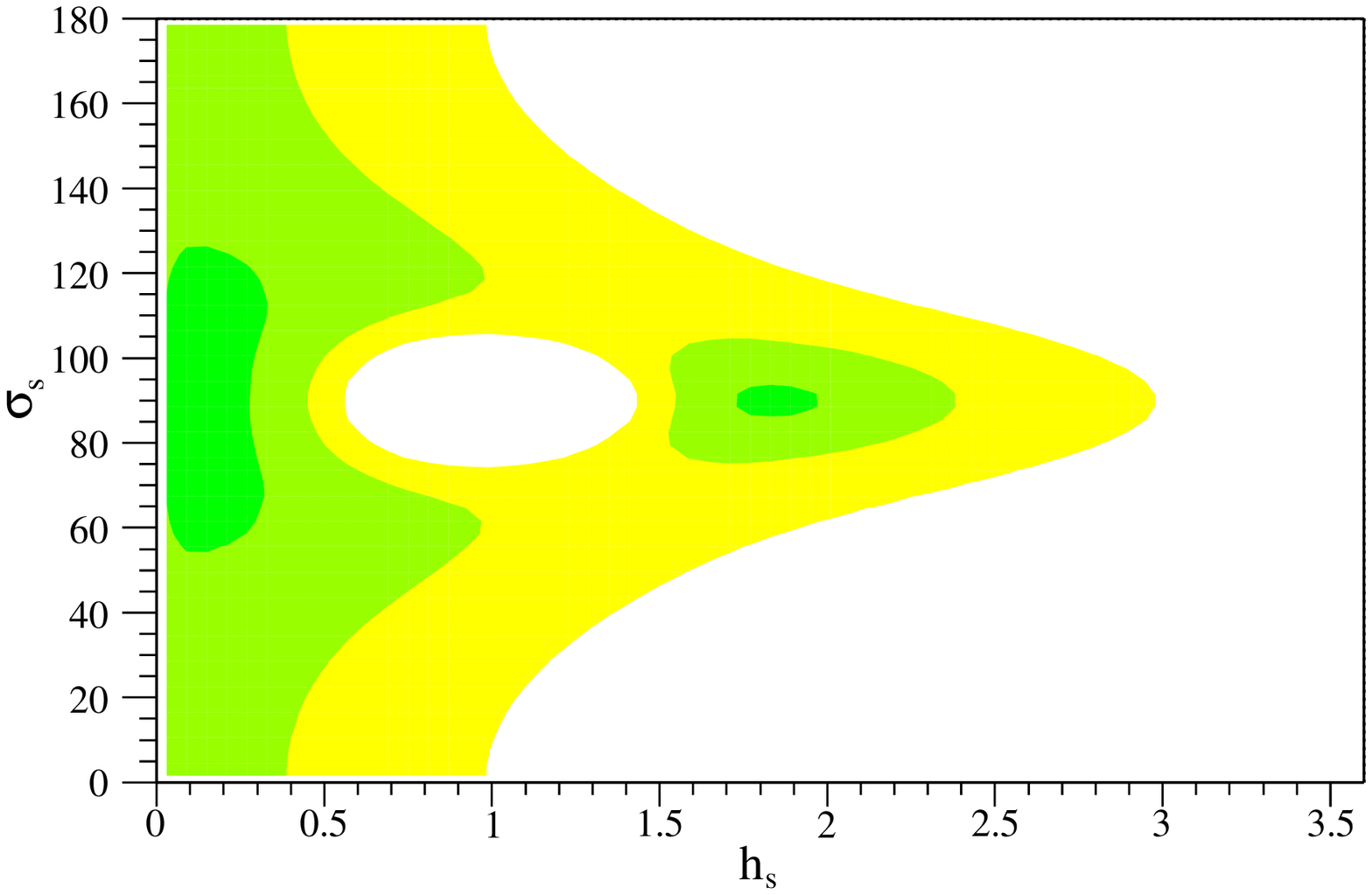}}
\caption{Allowed range for NMFV parameters
$h_s,\sigma_s$ using the data before (left) and after (right) the 
$\Delta M_s$ measurement. The dark/medium/light 
areas have CL $>$ 0.90, 0.32, and 0.05, respectively.
From hep-ph/0604112.}
\label{fighsss}
\end{figure}

\begin{figure}[!tb]
\begin{center}
\begin{tabular}{c c}
	\includegraphics[width=0.4\textwidth]{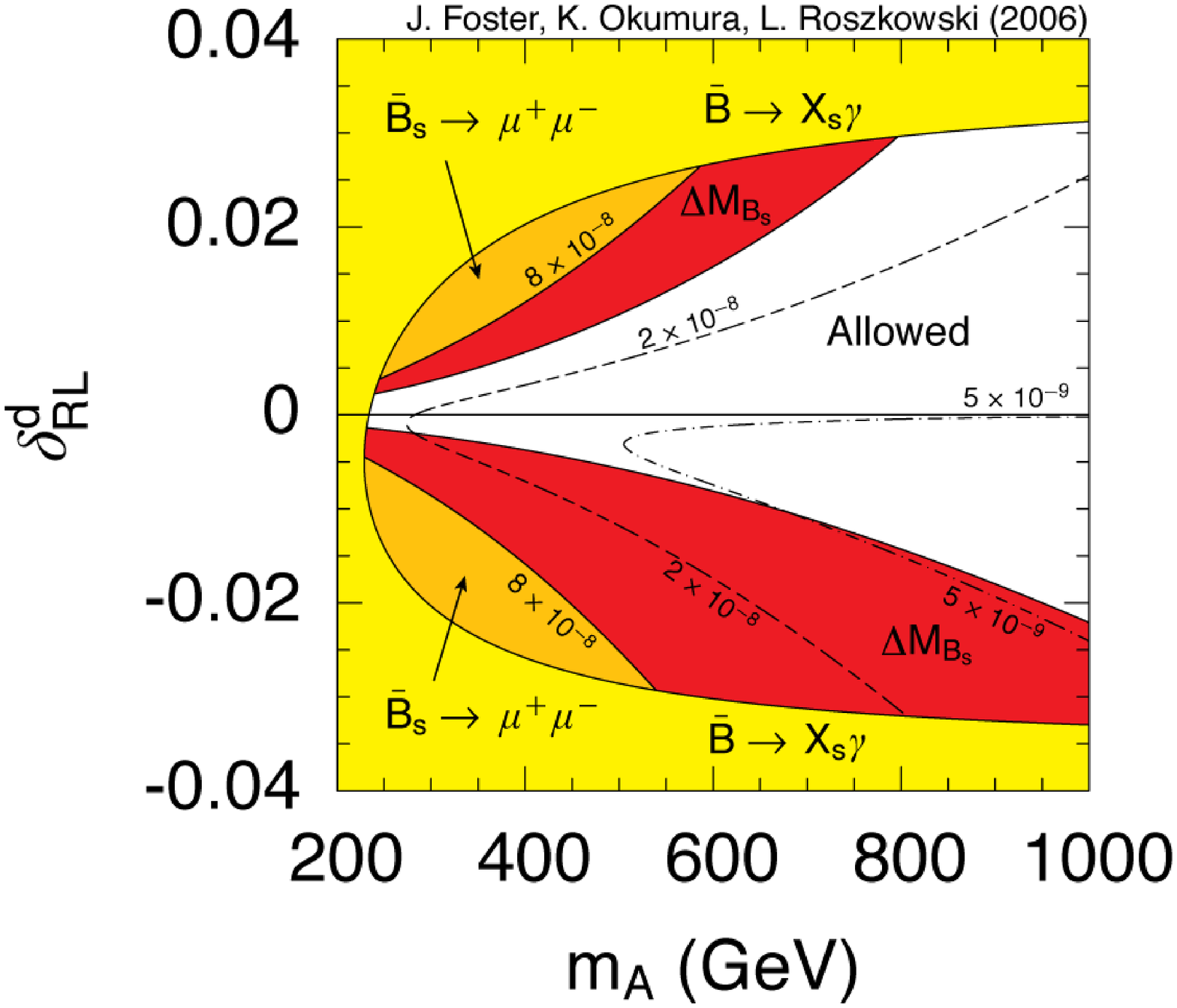}
&	\includegraphics[width=0.4\textwidth]{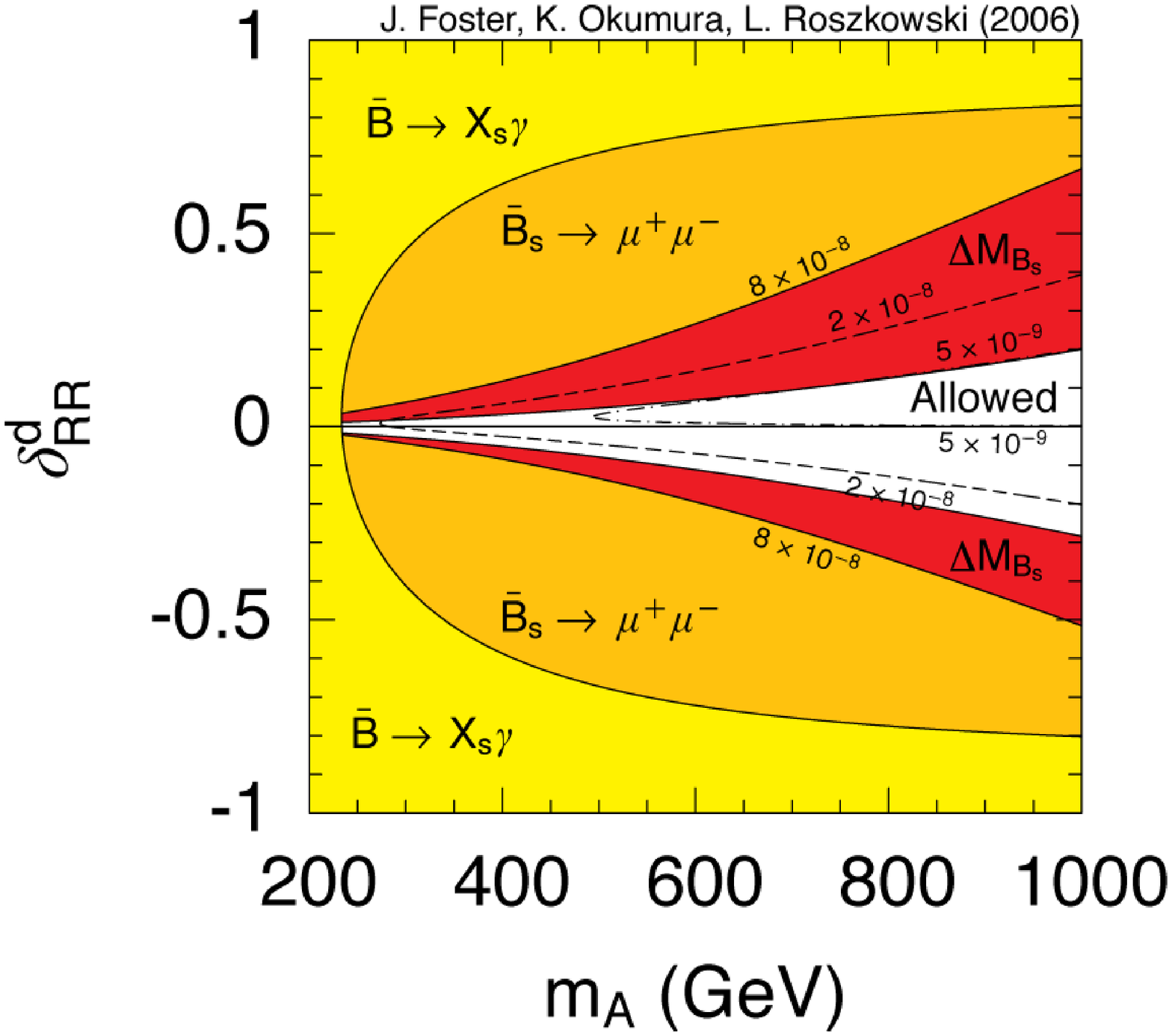}
\end{tabular}
\end{center}
\caption{
Contour plots showing the limits on the GFV supersymmetry parameters
$\delta^d_{RL}$ and $\delta^d_{RR}$ for varying $m_A$ 
and fixed tan$\,\beta = 40$, $\mu = -A_u = 500$ GeV,
and a common squark/gluino mass of 1 TeV. From hep-ph/0604121. 
}
\end{figure}

\section{The SciFi Channel}

Moving beyond the Standard Model,
let us contemplate the opening titles of a popular program 
on The SciFi Channel:
\begin{itemize}
\item The Cylons were created by man.
\item They evolved.
\item They rebelled.
\item There are many copies.
\item And they have a plan.
\end{itemize}
This is a perfect outline to review the history
and status of BSM theory.

\subsection{The models were created}
If you attended Moriond circa 1983, you recall that BSM theory
consisted of supersymmetry, grand unification, and technicolor.
The technical and phenomenological status of these models
was primitive. The BSM community was compact, and did not include
the even smaller detached cults of ``neutrino'' and ``particle-astro''
people.

\subsection{They evolved}
During the intervening 20+ years, there has been enormous
development in BSM theory. String theory took over the BSM
high ground, at first discouraging phenomenological progress,
but ultimately stimulating it with new ideas and powerful
technical insights. Supersymmetry models became much more
sophisticated, detailed and ambitious, creating a framework
with the potential to describe everything from Higgs to
unification to dark matter to inflation to baryogenesis.
Technicolor was badly mauled by electroweak precision data,
but revived with help from AdS/CFT and other technical
advances \cite{Sannino:2006wk}.

\subsection{They rebelled}
Despite this progress, the BSM community has been increasingly
unsettled. After 30 years, SUSY is still not discovered,
which is surprising given the golden opportunities from LEP,
the Tevatron, B physics, electric dipole moment measurements, etc.
Meanwhile the mysteries of flavor, as we have already seen above,
have gotten worse, compounded by the even deeper mystery of
how (or if) gravity couples to vacuum energy. 

Many theorists responded by moving in directions orthogonal
to the traditional main line of BSM development. There was
an explosion of model building invoking various scenarios
with extra dimensions. The extra dimensions could be
infinitely large but hidden, very large (10 fm to 0.1 mm),
large (1/TeV), or tiny (1/$10^{16}$ GeV) but 
warped \cite{Quiros:2006my}.
They have been invoked to solve the hierarchy problem,
break SUSY \cite{Shadmi:2006iw}, 
explain dark matter \cite{Flacke:2006tj}, 
explain some hierarchies of fermion masses \cite{Moreau:2006np,Frere:2004yu},
and even to explain the accelerating expansion of 
the universe \cite{Navarro:2006mq}.
None of these models are as robust and well-developed as
traditional SUSY, but in many cases they are compatible
and complementary with SUSY models, as well as with other frameworks
like Little Higgs \cite{Berezhiani:2005pb} or technicolor.

The most dramatic examples of BSM theory rebellion are
the Higgsless models and split-SUSY. The 
Higgsless models \cite{Simmons:2006iw}
use extra heavy gauge bosons, instead of a Higgs, to unitarize
the scattering of longitudinal $W$'s and $Z$'s. While generic
examples of such models are already ruled out by electroweak
precision data, they are a warning that Nature may have chosen
a more obscure path to electroweak symmetry breaking than we
have yet imagined. Split-SUSY enthusiasts
throw out the baby and play
with bathwater, creating models \cite{Arkani-Hamed:2004fb}
which explain dark matter and unification,
have nice flavor properties, but give up on explaining the origin
or stability of the electroweak scale. 

\subsection{There are many copies}

\begin{figure}
\begin{center}
\psfig{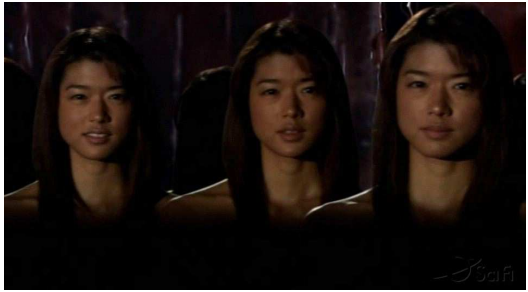}
\caption{With BSM models, as with Cylons, there are many copies.
\label{fig:sharon}}
\end{center}
\end{figure}

Despite wildly differing theoretical inputs and philosophies,
many BSM models end up with similar phenomenological outputs.
This is because nearly all of these models are trying to do
the same three things:
\begin{itemize}
\item Explain electroweak symmetry breaking.
\item Explain dark matter.
\item Avoid being ruled out by constraints from current 
data \cite{Marandella:2006ba}.
\end{itemize}
Thus most BSM models have a WIMP dark matter candidate;
other heavy exotics can decay to this WIMP, leading to the
prediction of dramatic missing energy signatures at colliders.
To avoid constraints from electroweak precision data, from
data on flavor violation, and other low energy precision
measurements, BSM models choose from three strategies:
\begin{itemize}
\item The new states are very heavy (multi-TeV).
\item There are conspiracies to cancel electroweak radiative
corrections and/or flavor violating effects.
\item A conserved parity requires all 
of the new particles to be pair-produced (suppressing
radiative corrections and providing a stable WIMP),
and the model is Minimal Flavor Violating or at least
NMFV.
\end{itemize}
The third strategy is the most attractive, leading to
a variety of SUSY models, Little Higgs with T-parity, and
Universal Extra Dimensions. It will be quite challenging
at the LHC to tell these models apart (see Figure 4).
This is true even
if we restrict to just SUSY models \cite{White:2006wh}. 

\subsection{And they have a plan}

The plan is not to replace the Standard Model. The plan
is rather to discover the larger more explanatory framework
in which the SM is embedded. I expect that discoveries
of the next ten years will teach us as much about the Standard
Model itself as what lies beyond it.

\section{The Big Picture}

\begin{figure}[t]
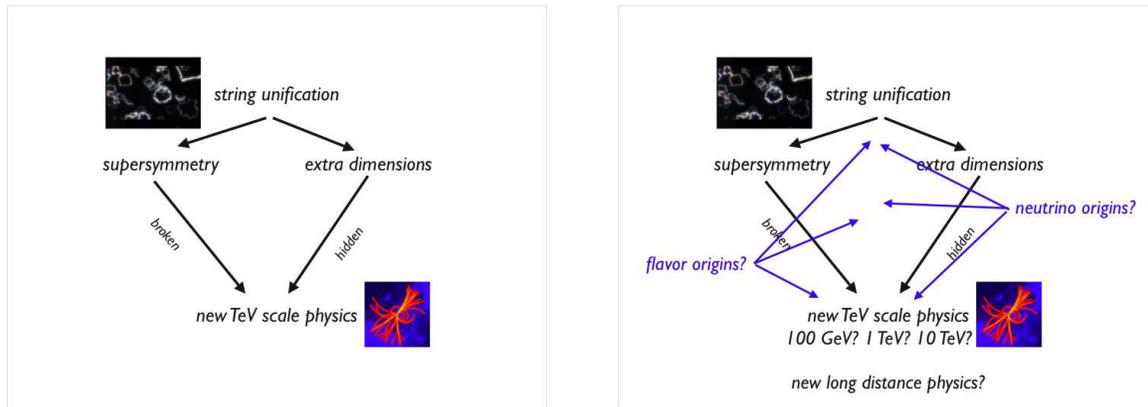

\centerline{\includegraphics[width=.5\columnwidth]{big1.epsf} \hfil
  \includegraphics[width=.5\columnwidth]{big2.epsf}}
\caption{The big picture of physics Beyond the Standard Model.}
\label{bigpicture}
\end{figure}

The big picture \cite{Lykken:2005up}
of BSM physics is illustrated in Figure 5.
The vertical direction represents energy scale; shown
are the ``TeV scale'' characterized by the new physics
responsible for electroweak symmetry breaking, and
the ``string unification'' scale, defined as the scale
where one begins to have a unified description
of quantum gravity with the gauge interactions of the SM.
We do not know the scale of unification even to within
an order of mangitude, nor do we know to what extent
it involves gauge coupling unification, grand unification,
flavor unification, or superstrings. Our best guess is that
some combination of all these elements is involved.

If unification occurs in any form, then there must be highly
sophisticated dynamical mechanisms which convert the simple
unified theory at ultra high energies into the messy junk
that we observe in experiments. These are shown in the figure
as the mechanisms that break supersymmetry and hide extra
dimensions. It has also been suggested that the messiness
we observe at accessible energies is due mostly to initial
conditions, selecting from a vast ``landscape'' of possible
vacua. This possibility involves strong cosmological assumptions
whose plausibility I find hard to evaluate.

Even with the caveats just mentioned, the left panel of
Figure 5 displays an elegant picture that is 
well-motivated and widely believed. However the right
panel is a more honest depiction of our current state
of ignorance. We know that neutrinos have mass, but
the physics responsible for this may lurk anywhere
in a 15 order of magnitude energy range. This is also
true {\it a fortiori} for the origin \cite{Morisi:2006tw}
of the complicated flavor
structure of the Standard Model.

To see better where we stand, we can break down our
ignorance about one subject, neutrinos, into a set
of concrete questions \cite{Petcov:2006pc}:
\begin{itemize}
\item What energy scales and symmetries are involved
in the orgin of the PMNS masses and mixings?
\item How are these related to the CKM matrix?
\item Are the masses Dirac or Majorana?
\item Is the mass hierarchy normal, inverted or quasi-degenerate?
\item What is the absolute scale of neutrino masses?
\item Are there light sterile neutrinos? If so, are they eV of keV?
\item What is the relation to dark matter?
\item What is the value of $\theta_{13}$?
\item Is there $CP$ violation in the lepton sector? If so,
are the $CP$ phases Dirac or Majorana or both?
\item Is there lepton flavor violation apart from Majorana masses?
\item How are neutrinos related to leptogenesis/baryogenesis?
\end{itemize}

The prospects for neutrino physics in the next decade are very
promising. One can imagine, through a combination of proposed
accelerator and non-accelerator based experiments, piecing
together an exciting story of neutrino origins. For example,
such a story could tie supersymmetry at the LHC to observations
of novel lepton flavor violation plus neutrinoless double beta decay,
and a discovery of an inverted neutrino mass hierarchy. In this
case we could develop a compelling theory of leptogenesis.
Many similar stories were discussed at 
Moriond \cite{Petcov:2006pc}{}$^-$\cite{Underwood:2006xs}.

\section{TeV Cosmology}

The WMAP three year results \cite{Spergel:2006hy} arrived during
the week of Moriond. These add yet more independent evidence
for the reality of dark matter. The rival MOND explanation
of galaxy rotation curves \cite{Navarro:2006mq} appears to
be under serious attack from both small scale data (dwarf galaxies)
and large scale data (galaxy cluster collisions).
WIMPS and axions are both well-motivated dark matter
candidates, and both are getting constrained 
by direct searches \cite{Sikivie:2006ir,Leclercq:2006zu}.
Of course direct and indirect searches are affected by
astrophysical uncertainties
about the local density of dark matter and how it is distributed
in the galaxy, so a positive signal is more informative
than a negative one.

The LHC collider provides a golden opportunity to 
manufacture and study WIMP dark matter in the laboratory. 
My guess is that dark matter will turn out to consist of
several different components, just as visible matter does,
and that a thermal relic WIMP will be an important part
of the story.

Still there are many challenges for understanding WIMP dark
matter. Neutralino dark matter from SUSY is the best understood
case, but here we know that the relic density estimates
are strongly dependent on details of the model.
Thus, for example, it is certainly not sufficient just
to scan over the CMSSM parameter space \cite{Cerdeno:2004zj}.
Other SUSY dark matter candidates need to be taken 
seriously \cite{West:2006tn}. Alternatives to SUSY, such
as Universal Extra Dimensions and Little Higgs with T parity,
are still baby models which will become more sophisticated
over time; their WIMP relic density estimates will evolve
accordingly.

One of our great ambitions for the LHC/ILC era should be
to uncover the hard details of TeV cosmology. 
At present, the earliest clear signpost of our cosmological
history is from the MeV scale, the time of primordial
big bang nucleosynthesis (BBN). Particle physics data
implies that a quark deconfinement transition happened
at higher temperatures, around 170 MeV. We expect to greatly
increase our knowledge of this transition during the next decade,
but not necessarily to obtain any strong links to cosmology.
Astrophysical data gives us hints
about a much earlier period of primordial inflation, but these hints are
clouded and ambiguous, much as are the hints of gauge/grand/string
unification in particle physics.

We strongly suspect that there was some kind of phase
transition at a temperature around 100 GeV, associated with
electroweak symmetry breaking. Combining results from LHC, ILC and other
experiments, we have a good chance of pinning down the
details of this transition, and its relation to 
baryogenesis/leptogenesis \cite{Huber:2006ik}.
Similarly, we suspect that at least one thermal relic
stable WIMP froze out during this same era of TeV cosmology.
Combining results from LHC, ILC and other
experiments on the nature of this WIMP and its interactions,
we should be able to compute its relic density under a
a variety of cosmological assumptions. As happened with BBN,
a successful linkup between these advances from particle
physics and those from dark matter searches and cosmological 
data could provide a sturdy TeV signpost for cosmology. 

Such a breakthrough may be essential for pushing on to
understand inflation and other features of our earliest
cosmological development. This necessity is illustrated
by noncanonical cosmological histories like the ``Slinky'',
which are nearly canonical from BBN time on, but wildly
different at earlier times \cite{Barenboim:2006iu,Barenboim:2006rx}.

Particle physics may also play a crucial role in
unraveling the mystery of ``dark energy''. This depends, however,
on what is the actual source of the current accelerating expansion.
There seem to be four general possibilities:
\begin{itemize}
\item The Friedmann-Robertson-Walker approximation is breaking 
down \cite{Kolb:2005ze}.
\item The Friedmann equation is modified due to extra dimensions
or modified gravity \cite{Navarro:2006mq}.
\item There is a tiny cosmological constant.
\item There is a dynamically evolving quintessence field.
\end{itemize}
In every case but the first it seems that particle physics input
will be essential.

\section{The LHC Era}
\begin{center}
``Never trust a theorist.''
\end{center}
\hspace*{9cm}\emph{-- Samuel C.C. Ting}

At Moriond there was palpable excitement about the advent
next year of the LHC. Those of us who need to abide by
U.S. Dept. of Energy travel rules have already made our
reservations to attend Moriond 2009, where we expect the
first LHC discoveries to be announced \cite{Lassila-Perini:2006sf}.
What will be found is anybody's guess, and if history is any guide,
even the most enlightened theorists will benefit from a
few sharp jolts of reality. One thing that we do know for sure
is that the LHC will open a window on a new world, even for
Standard Model physics. Understanding the Standard Model at
14 TeV is our most immediate challenge \cite{Gianotti:2005fm},
and one for which
close interaction between theorists and experimentalists
will be required.

\section*{Acknowledgments}
I would like to thank Jean Tran Thanh Van for 40 years
of leadership of the Rencontres de Moriond. I am grateful to
all of the Moriond organizers and staff for producing a
successful and enjoyable conference. Fermilab is operated
by Universities Research Association Inc. under Contract
No. DE-AC02-76CHO3000 with the U.S. Department of Energy.

\end{document}